# A Color Intensity Invariant Low Level Feature Optimization Framework for Image Quality Assessment

Navaneeth K. Kottayil · Irene Cheng · Frederic Dufaux · Anup Basu

**Abstract** Image Quality Assessment (IQA) algorithms evaluate the perceptual quality of an image using evaluation scores that assess the similarity or difference between two images. We propose a new low level feature based IQA technique, which applies filter-bank decomposition and center-surround methodology. Differing from existing methods, our model incorporates color intensity adaptation and frequency scaling optimization at each filter-bank level and spatial orientation to extract and enhance perceptually significant features. Our computational model exploits the concept of object detection and encapsulates characteristics proposed in other IQA algorithms in a unified architecture. We also propose a systematic approach to review the evolution of IQA algorithms using unbiased test datasets, instead of looking at individual scores in isolation. Experimental results demonstrate the feasibility of our approach.

## 1 Introduction

Affordable high quality imaging devices nowadays has led to advanced processing techniques and algorithms that make images easier to capture, store and transmit. However, these processing operations frequently introduce undesireable alterations to the original image that may not be visually pleasing. Too much compression, loss of contrast and color blurring are a few examples of such alterations. Quantitative and qualitative analysis of the resulting perceptual impact, generally known as image quality assessment (IQA), has been a research focus for decades. Robust image quality evaluation techniques have a wide range of applications, which include design and optimization along the image acquisition, processing and transmission pipeline. We focus on the factors causing degradation of image quality and adopt the Full Reference IQA approach in our experiments. Different IQA approaches can be found in the literature: explicit modeling of the HVS, use of structures in images, use of hand-crafted features (e.g., SIFT and HOG), or a combination of the above. In contrast, we present a state-of-the-art object detection based framework of convolutional neural networks, exploit the underlying concept to simulate the complex behavior of the Human Visual System (HVS), and apply the computed image feature representation for comparison of quality. The contributions of this paper include: (a) introduction of a new low level feature based IQA, (b) introduction of a new center-surround methodology taking visual saliency into consideration, (c) introduction of a new frequency scaling to optimize the extracted features, and (d) comparison with state-of-the-art methods to demonstrate better performance of our proposed method.

## 2 Motivation and Proposed Architecture

Based on earlier studies [25], it is believed that structural similarity plays a major role in perceptual quality assessment. Additionally, there are findings ( [11] and [30]) demonstrating that low level features can influence visual perception and can model how humans assess image quality. Recent theories ( [16] and [27]) also claim that the HVS in fact uses multiple strategies to

Navaneeth K. Kottayil, Irene Cheng and Anup Basu
University of Alberta
E-mail: kamballu;locheng;basu@ualberta.ca

Frederic Dufaux
Telecom ParisTech
E-mail: frederic.dufaux@telecom-paristech.fr

detect differences in images. The first mechanism dominates when the distortions are not easily visible. Here the visual system seems to employ a detection based strategy, that most of the IQA algorithms, e.g., [6], apply to explicitly model how the HVS detects quality. The second mechanism becomes effective when the distortions are supra-threshold or visible. Here the quality is determined mostly by the perceptual ability to recognize the image content. Visual saliency is another research approach to study IQA. Zhang et al. claimed that differences in visual salience maps can be used as a predictor of visual quality [29]. Another study [17] claims that SSIM can be further simplified by just considering gradient magnitude and choosing a better pooling strategy. The fundamental theory however is still based on the concept of structural similarity. An alternative method that has been successful is IFS [4] that tries to compute differences in terms of luminance and features extracted using Independent Component Analysis (ICA). ICA is also used to model the color mechanism of the HVS. Luminance distortion is calculated separately and combined with the ICA outcome to generate a final quality score.

We realized that all these theories can be accommodated into a single low level feature based model. Thus, we exploit an architecture that encapsulates the concept of object detection [14], and explore a feature optimization strategy to deliver a more efficient IQA framework. Current systems often capture edge like structures as features using natural images in the training set. The learnt filters are likely very similar to the Gabor filters that can be used to model the HVS [5]. In this type of model, high level features are generated from the input image. The generated feature representations are used for the object detection process. There are multiple processing layers, with each layer consisting of a filter-bank stage, a non-linearity stage and a feature pooling stage.

Instead of multiple layers, we only need one layer in our framework for generating low level features from the image. Our low level feature generation approach based on a single-layer object detection architecture is especially effective for IQA because of the following advantages: 1) the generated features capture structural information [25] of the image, 2) our approach, with a new frequency scaling technique, can explicitly model the HVS and account for the first mechanism to detect not easily visible distortions [16], 3) our approach captures the low level features that can be used for object detection accounting for the second mechanism to detect supra-threshold distortions [16], 4) the extracted low level features can influence visual perception based on earlier findings [11], 5) visual saliency is incorporated in our framework by using our new center-surround processing step.

## 3 Computational Model

Our model contains four major components: Filter-bank decomposition, Feature distribution normalization, Spatial frequency scaling and Neighborhood pooling. The evaluation scores of two images can then be compared to assess their perceptual similarity or difference.

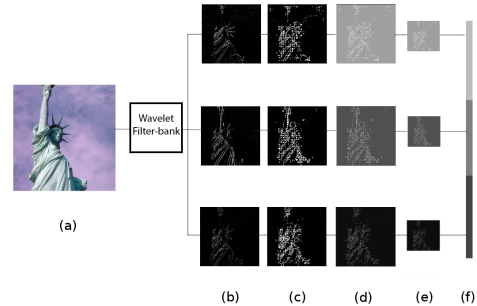

**Fig. 1** Outputs at various stages (a) The original image, (b) Three filter outputs, (c) Results after blockwise normalization, (d) Schematic representation of frequency scaling, (e) Pooling, (f) Pictorial representation of the final feature.

### 3.1 Pre-processing and filter-bank decomposition

An image is first converted into the CIELab color space generating the luma (L) and chroma (a and b) components. We adopt the CIELab color space for its larger gamut values to approximate human vision, exceeding those supported by the RGB and CMYK models. Each of the Lab components is then decomposed by a seven level wavelet, mimicking the frequency characteristics of the HVS [9]. We use BIOR 1.5 wavelets for this decomposition (Fig. 1 (b)). The output of these wavelet filters are represented by "feature maps." Note that we use a custom wavelet decomposition, rather than a trained set of filter weights described in the original model [14], the performance of which can be limited by the training set. A custom approach better simulates HVS characteristics like brightness induction and other frequency dependent ones [19].

### 3.2 Two-tier feature distribution normalization

The saliency of an image is dictated by the viewer's perceptual power to discriminate between the center and surround along with the relative distribution of features in the target region [2]. Motivated by this "center-surround" processing in biological vision, a two-tier normalization process is incorporated in our model. We perform patch-wise divisive normalization and subtractive normalization at each level of wavelet decomposition (feature map). A feature map is divided into patches of 13 x 13 with an overlap of 4 pixels between patches. The 5 x 5 center region within the patch is then processed. We use the surround to simulate the retinal eccentricity, beyond which the vision is blurred. Earlier studies [19] showed that this patch definition better simulates brightness induction of HVS.

3.2.1 Tier 1: Individual feature map normalization

A temporary subtractive normalization is first performed by subtracting the mean of the coefficients as shown in Eq. 1. $C_{s,o}(i,j)$ denotes the feature map at level $s$ and orientation $o$ of the wavelet decomposition for all pixels $(i,j)$ surrounding the current pixel in the 5 x 5 center block, and the mean of the coefficients is: $\overline{C_{s,o}(i,j)}$.

$$v_{s,o}(i,j) = C_{s,o}(i,j) - \overline{C_{s,o}(i,j)} \quad (1)$$

Let $\sigma_{v_{center}}$ be the standard deviation of the feature map values in a 5 x 5 neighborhood around the current pixel $(i,j)$ and $\sigma_{v_{surrounding}}$ be the standard deviation of the feature map values in the corresponding 13 x 13 neighborhood. We calculate the normalization factor $r$ for the divisive normalization as:

$$r = \begin{cases} \frac{\sigma_{v_{center}}}{\sigma_{v_{surrounding}}} & \text{if } \sigma_{v_{surrounding}} \neq 0 \\ \sigma_{v_{center}} & \text{if } \sigma_{v_{surrounding}} = 0 \end{cases} \quad (2)$$

Divisive normalization is essentially a decorrelation performed by dividing each pixel $v_{s,o}(i,j)$ of the 5 x 5 neighborhood by $r$.

$$y_{s,o}(i,j) = \frac{v_{s,o}(i,j)}{r} \quad (3)$$

The mean values are then added back to each feature map pixel $y_{s,o}(i,j)$. The temporary mean subtraction enhances only the variation in the visual information, without altering the mean value of the feature maps.

$$V'_{s,o}(i,j) = y_{s,o}(i,j) + \overline{C_{s,o}(i,j)} \quad (4)$$

3.2.2 Tier 2: Cross feature map normalization

After $V'_{s,o}(i,j)$ is computed for each feature map, the mean value $\overline{V'_{s,o}}$ across all feature maps is calculated. A subtractive normalization using this average is then performed across levels, i.e., level $s \in (2,7)$ and orientation $o$. $C'_{s,o}$ are the feature map values after the subtractive normalization.

$$C'_{s,o} = V'_{s,o} - \overline{V'_{s,o}} \quad (5)$$

The approximation feature map, i.e., level 1 (coarsest), is left unaltered in order to preserve the low frequency components that might have been lost on the final feature representation during the normalization and scaling processes.

Using this center-surround processing operation, we are able to simulate effects similar to lateral inhibition in the HVS, thus enhancing the regions of the image that have more variations. This has the added benefit of enhancing the visually salient regions of the image; e.g., global saliency computation in [13]. Salient regions have higher values in the final representation (Fig. 2).

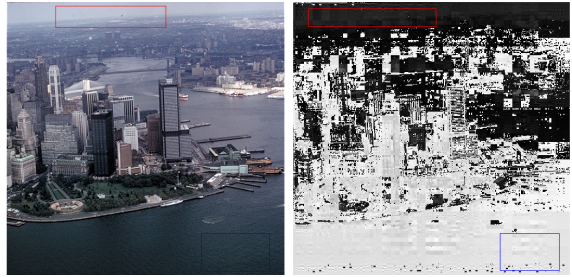

**Fig. 2** (Left) Original image. (Right) Result of reconstruction of the coefficients after normalization. The red rectangle (sky, far away buildings) denotes the region of low variation. The blue rectangle (city buildings, waves on the sea) is the region of high variation.

Liu and Heynderickx [18] improved the performance over SSIM [25] by scaling with the saliency term, following which they performed dissimilarity computation. We achieve the same effect in our method implicitly, by enhancing salient feature map values by our center-surround processing, before calculating perceptual distance.

3.3 Spatial frequency scaling optimization

The normalization step described in the last section allows us to capture the local features within each level (feature map) of the image. However, the HVS has different sensitivities to various levels and orientation of image edges. We observe that as the feature map detail increases (from coarse to fine) across levels, the corresponding variance associated with the feature maps decreases (Fig. 3). We examined the trends in thirty images and they display this characteristic collectively (Fig. 4). Feature maps at finer (higher) levels contain more detail and have smaller standard deviations. Motivated by this observation, we introduce a new spatial frequency scaling formulation to project the local frequencies at different levels to a global space.

3.3.1 Feature map scaling

We control the projection using the standard deviation of the feature map at each level and orientation, i.e., $\delta(s,o)$. We scale each of the feature maps by $\delta(s,o)$. The frequency scaled coefficients ($C_\delta(x,y)$) are:

$$C_{\delta(s,o)} = \delta(s,o) * C'_{s,o} \quad (6)$$

where the scale factor is:

$$\delta(s,o) = K_2/\sigma_{C_{s,o}} + K_1 \quad (7)$$

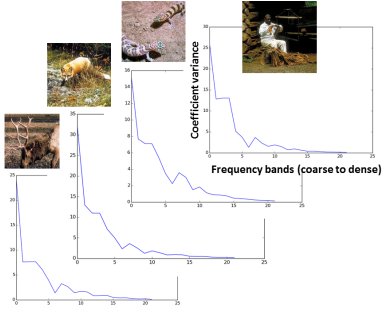

**Fig. 3** Example images illustrate the general trend of the coefficient variance associated with a feature map. It decreases when moving across levels.

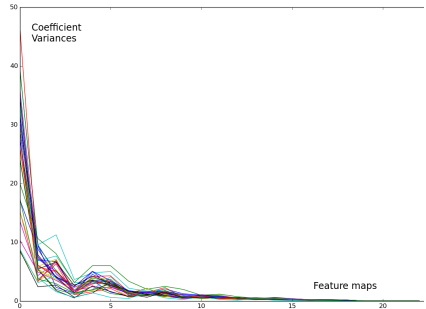

**Fig. 4** The consolidated result from thirty images also show that the variance coefficients in feature maps decrease when moving across levels.

where $s$ is level of feature map; $o$ orientation of the filter associated with horizontal, vertical and diagonal feature maps after wavelet decomposition; and $\sigma_{C_{s,o}}$ the standard deviation of the processed feature map at level $s$ and orientation $o$.

An optimization can be formulated by choosing the appropriate values of $K_1$ and $K_2$. For illustration, we plot the SSRC correlation scores in Fig. 5 using different pairs of $K_1$ and $K_2$ values. More detail about the correlation evaluation is presented in the results Section. The plot is the average result collected from 800 images. In our experiments, we set $K_1$ and $K_2$ to 31 and 3 respectively, which produced the optimal results. Coefficients of the frequency band and orientation that have higher scaling value $\delta(s,o)$ have a greater value in the final feature representation, corresponding to higher sensitivity to the HVS.

### 3.3.2 Color adaptation

It is believed that sensitivity of the HVS to scene content is derived from various stimulus modalities, including intensity, color, spatial and temporal (for dynamic scene only) features [7]. This is consistent with what we noticed from our datasets, where brightly colored images display a different perceptual sensitivity compared to the less vibrant images. To address this issue, we assess the chroma component of an image. If the color ratio $Cr > 0.25$ (Eq. 8), we adaptively modify the frequency scaling function of each frequency band by $(K_2/\sigma_{L_s} + K_2/\sigma_{C_{as,bs}} + K_1)$, where $\sigma_{C_a s}$ and $\sigma_{C_b s}$ are the standard deviations of the processed chroma feature maps at level $s$. $K_1$ acts as a decorrelation term for the color components of the image.

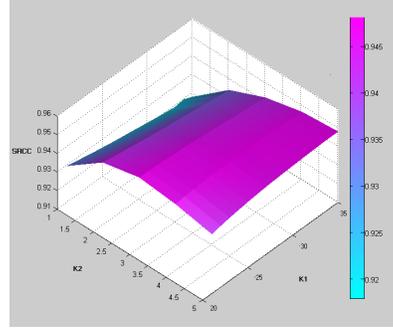

**Fig. 5** An optimization formulation can be obtained by adjusting the values of $K_1$ and $K_2$.

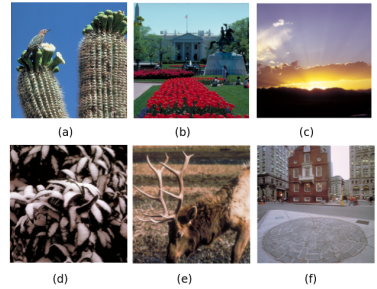

**Fig. 6** Colorful images (a),(b) and (c) are recognized by $Cr >= 0.25$. Example images with $Cr < 0.25$ are shown in (d),(e) and (f).

$$Cr = \sum_{s=1}^{n} \frac{\sigma_{as} + \sigma_{bs}}{(\sigma_{Ls})} \quad (8)$$

$\sigma_{Ls}$ denotes the standard deviation of the feature map of the luminance component at level $s$. $\sigma_{C_{as,bs}}$ is the product of $\sigma_{as}$ and $\sigma_{bs}$. The experimentally determined threshold of 0.25 was chosen to distinguish between colorful and normal images. A comparison of the two categories of images is shown in Fig. 6.

### 3.4 Neighborhood pooling

Aggregating multiple low level features over a small neighborhood can improve visual tasks [8]. In our model, the pooling component takes an $N \times M$ block in a processed feature map, divides it into $k \times k$ blocks and returns a single value equal to the maximum valued coefficient in the block. Thus, we reduce the dimensionality of the feature representation from $N \times M$ to $\frac{N}{k} \times \frac{M}{k}$. We choose $k$ as 3 for a window size of 3 x 3 and select the maximum of the feature map values in the block. The dimensionally reduced feature map $C_f(s,o)$ is computed as:

$$C_{f(s,o)} = pooling(C_\delta(s,o)) \qquad (9)$$

To create the final feature representation $f$ for an image, we apply concatenation on all the processed feature maps $C_{f(s,o)}$.

$$F = concatenate(C_{f(s,o)}) \qquad (10)$$

This is done for every level $s$ and orientation $o$ of the processed feature map $C_{f(s,o)}$.

3.5 Perceptual distance measurement

In order to compute the perceptual similarity between images, we use the $L1$ norm to compare the features between images.

$$e = \sum_{i=1}^{N} |F_1(i) - F_2(i)| \qquad (11)$$

$F_1$ and $F_2$ are the two features generated from Images 1 and 2 respectively; $N$ is the dimension of the features; and $e$ is the perceptual difference of Image 1 with reference to Image 2.

4 Experimental results

We tested our framework on four benchmark databases, which contain a set of original images, the degraded version and the perceptual quality scores, i.e, mean opinion scores (MOS). We generated quality scores on these images using our algorithm. For a fair comparison with other algorithms a logistic function was fitted to get a non-linear mapping from the objective scores to the subjective scores, following [22]. The comparison was based on Spearman rank order correlation coefficient (SRCC), Kendall rank-order correlation coefficient (KRCC) and Pearson linear correlation coefficient (PLCC). A good IQA is characterized by higher values for SRCC, KRCC and PLCC. Our implementation in python had an average time per image pair (1 reference and 1 distorted) of 3.06s for the LIVE dataset. This can be improved with C++ programming. Our computational complexity is similar to that of SSIM; one wavelet decomposition (FIR implementation O(N log(N))) followed by a scaling in windows and then on sub-band level. Since a wavelet transform is used, the total number of pixels in all the sub-bands remains the same as the original image, resulting in low overall complexity.

4.1 Performance comparison and analysis

The strengths of our model can be attributed to the local normalization and global scaling processes, which have advantages over traditional methods. These operations model the adaptation of the HVS by decorrelating elements in the feature maps along the axis of the wavelet basis, mimicking the processing in retinal ganglion cells [1] [10]. The end result is similar to PCA whitening (where the whitening operation makes the different components of PCA uncorrelated and of unit variance). The color intensity invariance feature of our model improves the IQA results by adaptively categorizing images at different brightness levels and analysing accordingly.

To test the performance, we run our IQA algorithm on the CSIQ [15], LIVE [12], TID2013 [20] and TID [21] databases. A comparison among different IQA algorithms on various datasets is given in [27], which we use to evaluate our algorithm. The algorithms that we compare with are VSI (Visual Saliency induced IQA) [29], PSNR, SSIM (Strutural similarity based IQA) [25], MS-SSIM (Multi scale SSIM) [26], VSNR [3], VIF (Visual information fidelity based IQA) [24], FSIM (Feature similarity index IQA) [30], IW-SSIM ( information weighted SSIM) [23], IFS (Independent feature detector IQA) [4], GSIM (Low level gradient similarity IQA) [17], MAD (Most Apparent Distortion) [16], GMSD ( Advanced SSIM based on gradient magnitude IQA ) [28]. The results are shown in Table 1.

Over the years, many IQA algorithms have been introduced. In order to evaluate the correlation between subjective scores and objective scores generated by IQA algorithms, statistical ranking methods like SRCC, KRCC and PLCC are used. However, what is the significant threshold in these rankings which truly reflects noticeable visual quality difference in the assessed images? Is it 0.01 or 0.05? In the VSI paper, the authors highlight the top two scores with a difference up to 0.05. In Table 1, we bolded the scores which are within 0.03 (half way between 0.01 and 0.05) of the maximum value. Our method has all bolded scores while others have at least one score not bolded. For comparison, if the threshold is reduced to 0.02, the proposed method only has one not bolded score while IFS and VSI have 3 and 6 respectively. One obvious reason for our consistent performance, as illustrated in Table 1, is that other algorithms work particularly well in one test database at the expense of another. For example, LIVE was released around 2005 by the authors/co-authors of [SSIM, MS-SSIM, VIF, FSIM, IW-SSIM and GMSD] (published in 2004, 2003, 2005, 20011, 2009 and 2013 respectively). The techniques described in these papers, following the concept of SSIM, all perform well in LIVE. Image structural content is an important factor for quality assessment, but an algorithm designed for one type of structure, e.g. edges, may not be effective on another. Be-sides, we observed that by tuning the parameters in an algorithm, the outcome may favor one test dataset over another. By adopting an optimal set of parameter values, as in our method and as explained in the IFS paper, a balance in quality across datasets can be achieved.

**Table 1** Comparison of performance on datasets

|  |  | Proposed | VSI [29] | PSNR | SSIM [25] | MS-SSIM [26] | VSNR [3] | VIF [24] | FSIM [30] | IW-SSIM [23] | IFS [4] | GSIM [17] | MAD [16] | GMSD [28] |
|---|---|---|---|---|---|---|---|---|---|---|---|---|---|---|
| LIVE (2005) | SRCC | **0.9430** | 0.9524 | 0.8756 | **0.9479** | **0.9513** | 0.9274 | **0.9636** | **0.9634** | 0.9567 | 0.9599 | 0.9554 | 0.9669 | 0.9600 |
|  | KRCC | **0.8200** | 0.8058 | 0.6865 | 0.7963 | 0.8045 | 0.7616 | **0.8282** | **0.8337** | 0.8175 | 0.8254 | 0.8131 | 0.8421 | - |
|  | PLCC | **0.9467** | 0.9482 | 0.8723 | **0.9449** | **0.9489** | 0.9231 | **0.9604** | **0.9597** | 0.9522 | 0.9586 | 0.9437 | 0.9674 | 0.9600 |
| TID (2008) | SRCC | **0.8820** | 0.8979 | 0.5531 | 0.7749 | 0.8542 | 0.7046 | 0.7491 | **0.8805** | 0.8559 | **0.8903** | 0.8554 | 0.8340 | **0.8910** |
|  | KRCC | **0.6969** | 0.7123 | 0.4027 | 0.5768 | 0.6568 | 0.5340 | 0.5860 | **0.6946** | 0.6636 | **0.7009** | 0.6651 | 0.6445 | - |
|  | PLCC | **0.8883** | 0.8762 | 0.5734 | 0.7732 | 0.8451 | 0.6820 | 0.8084 | **0.8738** | 0.8579 | **0.8810** | 0.8462 | 0.8306 | **0.8710** |
| CSIQ (2010) | SRCC | **0.9432** | 0.6423 | 0.8057 | 0.8755 | 0.9132 | 0.8105 | 0.9194 | 0.9242 | 0.9212 | **0.9581** | 0.9126 | **0.9467** | **0.9560** |
|  | KRCC | **0.7879** | 0.7857 | 0.6078 | 0.6900 | 0.7386 | 0.6241 | 0.7532 | 0.7561 | 0.7522 | **0.8158** | 0.7403 | 0.797 | - |
|  | PLCC | **0.9395** | 0.9279 | 0.8000 | 0.8612 | 0.8991 | 0.8002 | **0.9278** | 0.9120 | 0.9144 | **0.9576** | 0.8979 | **0.9502** | **0.9540** |
| TID2013 | SRCC | **0.8829** | **0.8965** | 0.6394 | 0.6274 | 0.7851 | 0.6818 | 0.6769 | **0.8015** | 0.7779 | **0.8697** | 0.7846 | 0.7808 | - |
|  | KRCC | **0.6979** | **0.7183** | 0.4696 | 0.4554 | 0.6029 | 0.5084 | 0.5147 | **0.6289** | 0.5977 | **0.6785** | 0.6255 | 0.6035 | - |
|  | PLCC | **0.8890** | **0.9000** | 0.7017 | 0.6861 | 0.8334 | 0.7129 | 0.7720 | 0.8589 | 0.8319 | **0.8791** | 0.8267 | 0.8267 | - |

Among the state-of-the-art IQA techniques from 2005 to 2015 no one algorithm performs best for all datasets. GMSD does not show KRCC score and the test on TID2013 is missing. Thus, we exclude it from the comparison. Both IW-SSIM and MAD were published in 2009. While IW-SSIM outperforms MAD in TID, MAD is better in CSIQ and LIVE. Both algorithms work equally well in TID2013. While MAD works better in all four datasets than VIF (2005), IW-SSIM is not as good as VIF in the LIVE dataset. FSIM was published in 2011. Although it shows improvement over IW-SSIM in the TID2008/2013 datasets, MAD (2009) is better than FSIM in the CSIQ and LIVE datasets. VSI and IFS were published in 2014 and 2015 respectively. VSI shows the best results in the two TID datasets and IFS is better in CSIQ, but MAD is still the best in the LIVE dataset.

Since IQA research has advanced rapidly in recent years, new parameters have been introduced in the algorithms to accurately assess image content. Accordingly, small image datasets are expanded to increase the variety of image content. It can be seen that compared with TID2008, TID2013 has seven extra types of distortions adding up to a total of 24 types. In comparison, LIVE (Release 2) has only five distortion types. Since TID2008 can be treated as a subset of TID2013, we exclude the scores of TID2008 to avoid double counting. Also, as pointed out in the IFS paper, "independent component analysis can provide a good description for the receptive fields of neurons in the primary visual cortex which is the most important part of the HVS." Image contents vary and each image is composed of low level components which stimulate the HVS. SSIM-based techniques detect certain types of component successfully, e.g., edge structures in the LIVE dataset. However, there are other perceptual components, such as luminance and color, generated from different types of distortion which are not described in the LIVE dataset. Thus, evaluation based on the scores in LIVE does not truly reflect potential distortions. Since TID2013 contains 3000 images and CSIQ (2010) contains 866 images, which are far more than other datasets, and they were created more recently, we use them as benchmark datasets for evaluating IQA algorithms.

|  |  | proposed | 2014 VSI | 2015 IFS | 2009 MAD | 2011 FSIM |
|---|---|---|---|---|---|---|
| CSIQ 2010 | SRCC | 0.9432 | 0.6423 | 0.9581 | 0.9467 | 0.9242 |
|  | KRCC | 0.7879 | 0.7859 | 0.8158 | 0.7970 | 0.7561 |
|  | PLCC | 0.9395 | 0.9279 | 0.9576 | 0.9502 | 0.9120 |
|  | avg | 0.8902 | 0.7854 | 0.9105 | 0.8980 | 0.8641 |
| TID2013 2013 | SRCC | 0.8829 | 0.8965 | 0.8697 | 0.7808 | 0.8015 |
|  | KRCC | 0.6979 | 0.7183 | 0.6785 | 0.6035 | 0.6289 |
|  | PLCC | 0.8890 | 0.9000 | 0.8791 | 0.8267 | 0.8589 |
|  | avg | 0.8233 | 0.8383 | 0.8091 | 0.7370 | 0.7631 |
| average |  | 0.8567 | 0.8118 | 0.8598 | 0.8175 | 0.8136 |

**Fig. 7** Our proposed method achieves a more consistent performance across datasets, while other methods have better scores in one at the expense of another dataset.

In a real-world application, it is not possible to predict what type or what combination of distortion(s) will occur in the processing, transmission and rendering pipeline, and accordingly select the best performing algorithm. Instead of comparing scores for 24 distortion types individually, it is practical to examine the overall (average) performance of an algorithm. Based on the test datasets CSIQ and TID2013, the 5 best performing algorithms with average score over 0.8 are MAD, VSI, IFS, FSIM and our proposed method. The average scores are shown in Fig. 7. Note that IFS and VSI have the best performance in the CSIQ and TID2013 datasets respectively, but at the expense of the other dataset. FSIM, IFS and MAD have a high difference in average score of more than 0.1 between the two datasets. Our algorithm has a high average and achieves more consistent performance in both datasets. Our method shows an improvement compared to low level feature based IQA FSIM [30]. The inclusion of two-tier normal-ization, optimized frequency scaling and color adaptation, attributes to this improvement.

**Table 2** Performance of using a single window compared with using a center-surround analysis

|      |      | Center-Surround | 3x3 | 5x5 | 7x7 |
|------|------|--------|--------|--------|--------|
| CSIQ | SRCC | **0.9432** | 0.9369 | 0.9346 | 0.9292 |
|      | KRCC | **0.7879** | 0.7792 | 0.7752 | 0.7662 |
|      | PLCC | **0.9395** | 0.9212 | 0.9183 | 0.9124 |
|      | RMSE | **0.0902** | 0.1026 | 0.1044 | 0.1079 |

To-date, there has been no one single IQA algorithm which outperforms others for all distortion types and in all benchmark test datasets. Our contribution lies in proposing a more consistent technique to assess image quality based on a systematic approach to review the evolution of IQA algorithms using unbiased test data, instead of following the traditional method to look at individual scores in isolation. The scatter plots of the scores generated by our metric against user subjective scores are shown in Fig. 8, which illustrates the consistency between our metric scores and the user scores. In order to illustrate the advantage of using center-surround in the normalization process, we computed the scores using only a simple window of size 3x3, 5x5 and 7x7 respectively. The results are shown in Table 2. The lower performance of single window compared with center-surround agrees with the finding that the saliency of an image is influenced by the relative distribution of the image features in the center and surround [2].

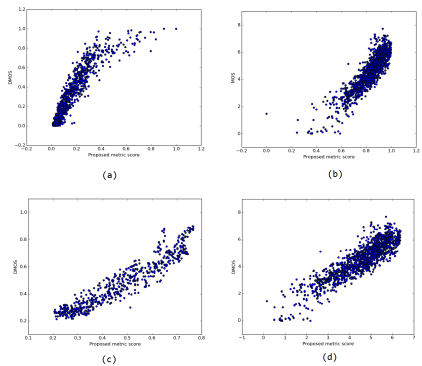

**Fig. 8** Scatter plots of our metric scores vs subjective scores on dataset (a) CSIQ, (b) TID, (c) LIVE and (d) TID2013.

An IQA algorithm often works well in one dataset but not in another. This is similar to machine learning algorithms whose performance depend on the training set. Depending on how the controlling parameters of an IQA algorithm are tuned, e.g., tuned with the CSIQ dataset, test images sharing similar characteristics with the training dataset will perform better using that algorithm. In order to illustrate this point, we fine-tuned the parameters $K_1$ and $K_2$ using the CSIQ dataset only. The optimal values were 33 and 5 respectively, leading to an increase of SSRC from 0.943 to 0.948 (Fig. 9), which is higher than the value 0.946 of MAD Fig. 7. However, the fine-tuned parameter values lead to a decrease of SSRC from 0.882 to 0.869 in the TID dataset. Until there is an IQA algorithm which can adaptively adjust the controlling parameters at the image level, it is difficult to have one algorithm that outperforms others in all image datasets. Given an arbitrary image, without knowing what training characteristics it is associated with, our method guarantees a balanced assessment.

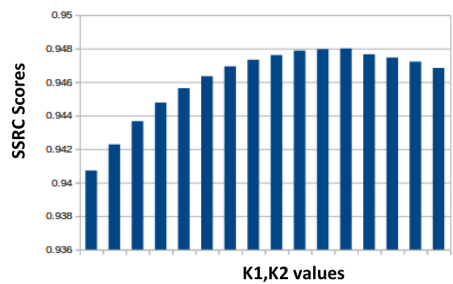

**Fig. 9** An IQA algorithm can perform better if fine-tuned based on a given database, but the same tuned parameter values are unlikely optimal when applying in another database. For example, an improved SSRC score of 0.948 was obtained when $K_1 = 33$ and $K_2 = 5$.

## 5 Conclusion

We introduced a new low level feature based IQA framework integrating many important characteristics: two-tier feature distribution normalization, frequency scaling optimization, and color adaptation. We exploit the latest object detection based architecture and incorporate our own normalization method and new frequency scaling functions. The resulting framework generates a concatenated feature vector for an image that captures the perceptual factors influencing visual quality. The difference in feature vectors between the reference image and the target image is computed as the perceptual difference between the two images. Tests performed on public datasets demonstrate that our proposed IQA has better consistent performance across databases. Furthermore, we propose a systematic approach to review the evolution of IQA algorithms using unbiased test datasets, instead of following the traditional method of looking at individual scores in isolation. In future work we will examine more complex pooling techniques and perceptual distance metrics. More importantly, we will examine our normalization and scaling techniques to adaptively compute the parameter values at image level.